\documentclass{elsart}
\usepackage{color}

\usepackage{epsfig}

\usepackage{amsmath}

\begin{document}

\begin{frontmatter}



\title{The rapidity structure of Mach cones and other large angle correlations in heavy-ion collisions }

\author[JYV,HEL]{Thorsten Renk,}
\author[McGill,Duke]{and J\"{o}rg Ruppert,}

\address[JYV]{Department of Physics, PO Box 35 FIN-40014 University of Jyv\"askyl\"a, Finland}
\address[HEL]{Helsinki Institut of Physics, PO Box 64 FIN-00014, University of Helsinki, Finland}
\address[McGill]{Department of Physics, McGill University,
Montreal, QC, Canada, H3A 2T8 }
\address[Duke]{Department of Physics, Duke University,
             Durham, USA, NC 27708-0305}

\begin{abstract}
The pattern of angular correlations of hadrons with a (semi-)hard trigger hadron in heavy-ion collisions has attracted considerable interest. In particular, unexpected large angle structures on the away side (opposite to the trigger) have been found. Several explanations have been brought forward, among them Mach shockwaves and Cherenkov radiation. Most of these scenarios are characterized by radial symmetry around the parton axis, thus angular correlations also determine the rapidity dependence of the correlation. If the observed correlations are remnants of an away side parton after interaction with the medium created in the collision, pQCD allows to calculate the distribution $P(y)$ of the away side partons in rapidity.
 The measured correlation then arises as a folding of $P(y)$ and the rapidity structure of the correlation taking into account the detector acceptance. This places non-trivial and rather stringent constraints on the underlying scenario. We investigate these dependences and demonstrate that Mach shockwaves survive this folding procedure well whereas Cherenkov radiation scenarios face new challenges.
\end{abstract}

\begin{keyword}

\PACS 
\end{keyword}
\end{frontmatter}

% main text

\label{}

\section{Introduction}

Announcements have recently been made by the four detector collaborations at RHIC \cite{RHIC-QGP} that a new state of matter, distinct from ordinary
hadronic matter has been created in ultrarelativistic heavy-ion collisions (URHIC). This state of mattter has shown a number of unexpected properties already. Recently, measurements of two-particle correlations involving one hard trigger particle have shown a surprising
splitting of the away side peak for all centralities but peripheral collisions, qualitatively very different from
a broadened away side peak observed in p-p or d-Au collisions \cite{PHENIX-2pc}. Interpretations in terms of energy
lost to propagating colourless \cite{Shuryak,Stoecker} and coloured \cite{Wake} sound modes have been suggested 
for this phenomenon. A comparison with data using a realistic trigger simulation has been performed in \cite{Mach}. As an alternative mechanism to generate such large angle correlations, Cherenkov radiation from the away side parton has been suggested \cite{Cherenkov}.

In hydrodynamical models for Mach shocks investigated so far \cite{Shuryak,Chaudhouri}, the simplifying assumption has been made that instead of considering the full three dimensional propagation of the Mach cone through the medium, a boost-invariant 'Mach-wedge' is simulated as an approximation of the situation close to midrapidity. Likewise, in scenarios where large angles arise directly from induced in-medium radiation \cite{Cherenkov,Vitev}, the angular structure is depicted for a single parton. In contrast, in experiment only the rapidity $y_{trig}$ of the trigger hadron is constrained to be inside the acceptance experimentally. This does however not determine the rapidity $y$ of the away side parton. For this, the conditional probability $P(y)$ given the rapidity and momentum of the parton leading to the trigger hadron can be obtained from pQCD. 
The measured correlation then results from an integral over all possible $y$ weighted with $P(y)$.

However, most models of jet energy loss would (without considering further interaction of modes excited in the medium) result in a correlation signal that exhibits angular symmetry with respect to the away side parton's axis. This means that for an away side parton at midrapidity, there is not only a correlation signal at large angle and zero rapidity but also a signal at small angle and large rapidity. Thus, the averaging over $P(y)$ will tend to smear the signal measured at midrapidity out towards smaller angles as compared to a simple midrapidity projection. It follows that the signal before averaging must be at even larger angles than a naive interpretation of the experimental data would suggest.

In the following, we will illustrate this point in greater detail. First, we derive the expression for $P(y)$ for the dominant reaction channels at typical trigger energies from LO pQCD. Then we show how the rapidity structure of the correlation signal arises from the angular information, both for scenarios in which the correlation propagates with the flowing medium and in those where it doesn't. We demonstrate within the full trigger simulation described in \cite{Mach} that a Mach shockwave is recovered in the detector acceptance sufficiently undistorted. Since no detailed comparison of a radiative large angle scenario with the data is available, we demonstrate how the large angle correlation relative to a single away-side jet must move to even larger angles than observed in experiment in such a description. 

\section{Rapidity distribution of away-side jets}

\label{Py}

The differential production cross-section of hard partons in A-A collisions can be obtained in leading order 
pQCD by folding the two particle production cross-sections $d\hat{\sigma}/d\hat{t}$ with the nuclear parton 
distribution functions $f_{i,j/A}$ (here we use \cite{NPDF}):
\begin{eqnarray}
\label{cross-y}
\frac{d^3\sigma^{AA \rightarrow k+l+X}}{dp_T^2dy_1dy_2}=\sum_{i,j} x_1 f_{i/A}(x_1,Q^2)x_2 f_{j/A}(x_2,Q^2) \frac{d\hat{\sigma}^{ij\rightarrow kl}}{d\hat{t}}\,\,.
%(\hat{s},\hat{t},\hat{u}).
 \end{eqnarray}

If the outgoing partons are at rapidities $y_1$ and $y_2$, $x_1$ and $x_2$
are determined by:

\begin{equation}
x_1 = \frac{p_T}{\sqrt{s}} \left[\exp(y_1) + \exp(y_2)\right] \quad \text{and}\quad
x_2 = \frac{p_T}{\sqrt{s}} \left[\exp(-y_1) + \exp(-y_2) \right]
\end{equation}

\begin{figure}[htb]
\epsfig{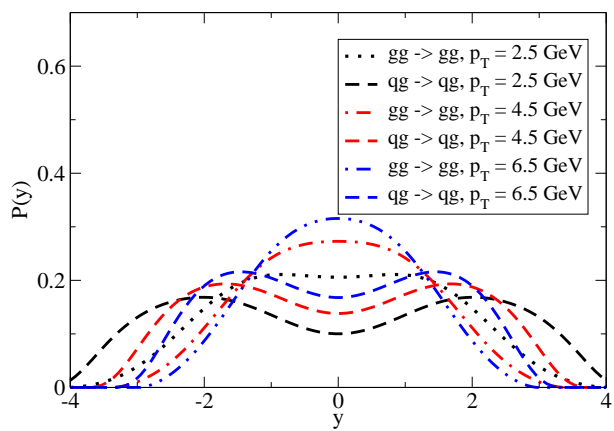}
\caption{\label{F-pQCD}Conditional probability $P(y)$ to find the away side parton at rapidity $y$ if the trigger parton is found at $y_{trig}=0$ calculated in LO pQCD for the dominant production channels in the low momentum regime as a function of different $p_T$.}
\end{figure}

The conditional probability distributions $P_{}(y)$ of producing an away-side parton  at rapidity $y$ can then be calculated from the normalized cross-section (Eq.~(\ref{cross-y})) given the trigger parton at $y_1=0$.
We show the normalized contributions of the two-dominant channels $gg \rightarrow gg$ 
and $gq \rightarrow gq$  to those conditional probabilities in Fig.~\ref{F-pQCD} for different $p_T$.
There is a significant production probability of away-side partonic jets in the range $\pm 2$ for lower $p_T$ which gets somewhat narrower as $p_T$ increases (note that the gluonic contribution dominates over the quark contributions in this momentum regime).

\section{Rapidity structure of the correlations}

The STAR and PHENIX experiments at RHIC measure the correlation
signal averaged over rapidity intervals of $\pm 1$ and $\pm 0.35$, respectively.
The measured signal must thus be understood as a superposition of 
individual two-jet production events in which the trigger condition was fulfilled. The trigger condition 
largely determines the weighted jet-production vertex distribution from which the near-side and 
away-side partons emerges. Since the trigger must be inside the acceptance (i.e. close to midrapidity), 
the away-side partons are be distributed in rapidity according
to the conditional probability distributions $P(y)$ derived in section \ref{Py}. 
We use a radiative energy loss formalism \cite{QuenchingWeights} to determine 
how much energy is deposited in the medium locally while the near-side and away-side partons traverse it. 
Characteristic $dE/d\tau$ distributions emerging from
the folding of the medium evolution with the jet-energy loss calculation are discussed in \cite{Mach}. 
Standard radiative energy loss calculations do not lead to angular structures  in the 
away-side jet's secondaries that could account for the observed large angle correlation, see e. g. \cite{Vitev}.

Several explanation have been brought forward to explain such large
angle correlations. It has been argued that those could emerge via the excitation of 
colorless \cite{Shuryak,Stoecker} or colorful \cite{Wake} sound modes  by the supersonically traveling away-side jet (excitation of 'Mach cones') or by the emission of Cerenkov-like gluon radiation \cite{Cherenkov} by the superluminally traveling jet in the nuclear medium. Colorful sound modes and Cerenkov-like gluon radiation could only contribute in a QGP phase and are only possible if a space-like longitudinal or
transverse dispersion relation is realized. This was pointed out first in \cite{Wake} and it has been shown in \cite{Cherenkov2} that those could emerge in a plasma if bound states are present. A space-like gluon dispersion relation does not emerge in HTL resummed calculations
of the longitudinal and transverse plasma modes which have time-like dispersion relations. 

First we focus on the excitations of Mach cones. We only give a brief overview 
since the details of the calculation employed here have already been discussed in \cite{Mach}. In addition, the rapidity structure is governed by rather general arguments for which details of the excitation are not relevant. We assume that a fraction $f$ of the locally lost energy of the away-side jet is transferred to a collective colorless mode with a linear dispersion relation $E=c_s p$, where $c_s$ is the speed of sound which is determined by the lattice EoS via $c_s^2=\partial p/\partial\epsilon$.
The evolution of the shock wave is tracked in the medium and the Mach angle
is determined via the averaged speed of sound during the evolution until freeze-out time as $\bar{c}_s=\int_{\tau_E}^{\tau} d\tau c_s(\tau)/(\tau-\tau_E)$. Finally the additional boost to hadrons due to the Mach shock wave is determined at freeze-out.  At momentum scales of $1$ GeV, well above typical temperature scales in the medium,  considerable contributions to the correlation signal are only expected where transverse flow and the Mach shock are aligned \cite{Mach2}. Freeze-out is then calculated using the Cooper-Frye formula. 

It has been pointed out in \cite{Stoecker2} that since the shock wave travels with $c_s$ in the local rest frame, the spatial position the of the shock front has to be determined by solving the characteristic
equation:
\begin{eqnarray}
\left.\frac{dz}{dt}\right|_{z=z(t)}=\left.\frac{u(z,R,t)+c_s(T(z,R,T))}{1+u(z,R,t)c_s(T(z,R,t))}\right|_{z=z(t)}.
\end{eqnarray}
In \cite{Stoecker2} it has been argued that this could destroy a Mach cone signal. This
statement seems to be based on a misinterpretation of the measurement: the observed correlation signal is not a Mach cone in position space but the resulting boost of hadrons in momentum space. Thus the position of the Mach cone at freeze-out is relevant only insofar as the longitudinal flow at $z_{\rm final}$ determines a longitudinal boost in momentum space. 
This means that a Mach cone in $\phi,y$-space is elongated significantly in $y$ direction by longitudinal flow. For instance for a Bjorken evolution this amounts to an elongation of a factor of about $1.5$ in rapidity.
%\textcolor{red}{
We refer to the effect described in this paragraph in the following as the 'effect of longitudinal flow' on the correlation signal where 'longitudinal' refers to the medium expansion parallel to the the beam direction and not along the direction of partonic jets.
%}

In order to compare with experiment, one has to fold the elongated Mach cone structure with the away-side jet distribution 
$P(y)$ and average the emerging correlation signal over the detector's rapidity acceptance.

This is very different from scenarios where no hydrodynamical mode is excited (e.g. %\textcolor{blue}{Cherenkov} \textcolor{red}{
gluon
%} 
radiation). Here, there is no reason to assume that the excited mode is moving with given velocity relative to the medium. If the emitted mode is not carried away by the flowing medium, the angular structure should translate directly into the observable signal. 

%\textcolor{red}{
This holds true
%} 
unless re-interaction with the medium is considered which then needs to be taken into account consistently also for the angular distribution. 
%\textcolor{red}{
We emphasize that the initial angular distribution of Cerenkov-like gluon radiation in a static medium faces these problems. The situation could be different if one takes into account how the initial emission structure Cerenkov-like gluons might be altered during the expansion of the medium where a changing index of refraction that depends on the 
medium evolution and the relative direction between jet and flow could 
introduce significant changes. We point out that since a theoretical analysis how this influences the predicted angular correlation signal in a Cerenkov-like gluon radiation scenario has not yet been performed, it is not clear
that this would eventually predict a considerable elongation of the 
cone in $y$ direction.
%}

\section{Results}

We calculate shockwave excitation in the dynamical evolution model framework as outlined above and compare with the $1 <p_T< 2.5$ GeV two-particle correlation signal for central collisions given a $2.5 <p_T< 4$ GeV  trigger averaged over the PHENIX detector's rapidity $|\eta|<0.35$ window. 
Since the fraction of the jet's lost energy that is transfered to the collective sound mode $f$ cannot yet be calculated from first principles, we treat $f$ as a parameter \cite{Mach}. 

\vspace{0.7cm}

\begin{figure}[htb]
\epsfig{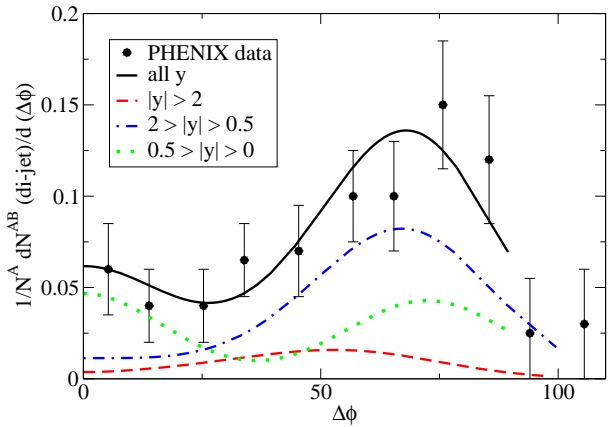}
\caption{\label{F-MachStd}Calculated 2-particle correlation on the away side for $|y|< 0.35$ and $1.0 < p_T < 2.5$ GeV. Indicated are also the partial contributions originating from away side partons going into different rapidity intervals given a trigger parton at midrapidity. }
\end{figure}

In Fig.~\ref{F-MachStd}  we show a comparison of our calculation with the PHENIX two-particle correlation data \cite{PHENIX-2pc} on the away-side for $f=0.75$\footnote{This value of $f$ differs somewhat from that one previously determined in \cite{Mach} were we used to make somewhat more simplistic assumptions about the rapidity structure of the source.}. Note that zero degrees is chosen such that it is opposite to the trigger, i.e. at the expected average position of the away side parton. We also show the relative contribution to this signal from 
Mach cones excited by away-side partons from different rapidity intervals. 
Contributions emerging from Mach cones from away-side jets produced at $|y|>2$ are 
suppressed since only part of the cone contributes in the detector's rapidity window $|y|<0.35$. The maximum of the $\phi$ distribution is shifted to lower angles $\phi\ll\phi_{\rm max}$, where $\phi_{\rm max}$ is the maximum of the calculated correlation signal for all $y$. Contributions emerging from Mach cones from away-side jets produced at $0.5<|y|<2$ contribute significantly at angle $\phi \sim \phi_{\rm max}$.
 The contribution at low angles $\phi<40$ degrees is dominated by contributions from the bow shock (i.e. the $(1-f)$ contribution to the deposited energy) emerging from away-side jets at $|y|<0.5$. Contributions of away side jets at $|y|<0.5$ are also important for the correlation signal around $\phi \sim 0$. This bow shock contribution falls almost completely out of the acceptance of the detector for away-side jets with $|y|>0.5$ as it is always very close to the rapidity of the away side parton.
 
\vspace{0.5cm}
\begin{figure}[htb]
\epsfig{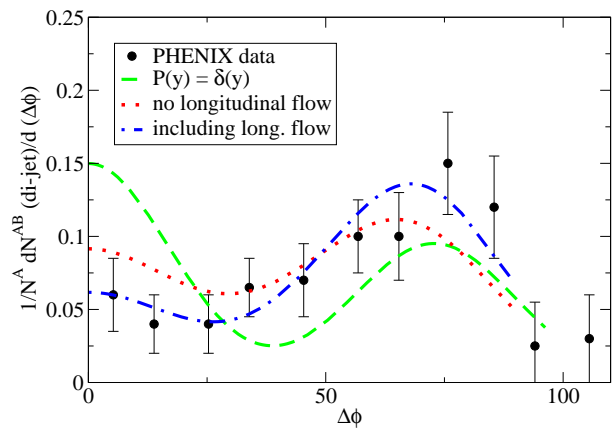}
\caption{\label{F-Elongation}Calculated 2-particle correlation under the assumption that a) the away side parton is always at midrapidity and the excited mode doesn't couple to flow (green) b) the excited mode doesn't couple to flow (red) and c) including realistic $P(y)$ and longitudinal flow effect. }
\end{figure}

Fig.~\ref{F-Elongation} illustrates what correlation signal would be expected if all away-side jets
were confined to mid-rapidity ($P(y)=\delta(y)$) and no flow would be present in comparison to the case where the  rapidity distribution of the away-side jet $P(y)$ as calculated in section (\ref{Py}) is appropriately taken into account.
The calculation is performed for the two-particle correlation signal as it is measured in the PHENIX detector's acceptance region 
$|\eta|<0.35$. The contribution for $\phi<40$ degrees from the bow-shock is much smaller for the wider $P(y)$ distribution of the away-side jets than for the narrow one, since the bow shock contribution from jets with $|y|>0.5$ falls almost completely out of the acceptance.
The large angle contribution is shifted to lower angles in comparison to the $P(y)=\delta(y)$ case, since the Mach shock cone of away-side jets with large $|y|$ contribute at lower $\phi<\phi_{\rm max, \delta(y)}$ in the detector acceptance interval $|\eta|<0.35$.  Here $\phi_{\rm max, \delta(y)}$ is the maximum
of the correlation signal at large angle if all away-side jets were confined to mid-rapidity.

Fig.~\ref{F-Elongation} also illustrates that the shift of the maximum of the correlation signal to lower
angles $\phi<\phi_{\rm max, \delta(y)}$ is more pronounced, if the elongation of the Mach cone in momentum space due to longitudinal flow would not have been taken into account. In addition the width of the two-particle correlation would increase. 
Such a correlation signal would no longer be in agreement with the PHENIX data. This argument can also be turned around: if no elongation would be present, the maximum of the emission angle from a single away-side jet at fixed $y$ has to be larger than $\phi_{\rm max}$ and the width considerably smaller than in the measured correlation signal. 
This is a strong constraint for gluon radiation scenarios, 
if the angular emission pattern of the radiated gluons is assumed to be directly translated in a two-particle hadronic correlation signal.

\section{Conclusions}
 
In this paper we have investigated the rapidity structure of two-particle correlation signals involving one hard trigger particle. We have shown that the
rapidity structure of the correlation signal arises because of two different reasons:
the away-side jets have a specific distribution in rapidity $P(y)$ that can be determined by pQCD since the near-side jet is almost centered at midrapidity\footnote{
%\textcolor{red}{
We mention that we checked that even if the near-side partonic jet is slightly (e.g. $\pm$ 0.3 units in rapidity)  off mid-rapidity the shape of the correlation signal (black line in Fig.2) remains essentially unaltered and no additional spread is introduced.}
%}
and the Mach shock fronts induced by those away-side jets in the medium are elongated along the rapidity axes by a longitudinal boost in momentum space due to longitudinal flow.

We have demonstrated that Mach cones as excited modes of the nuclear medium lead to the explanation of a correlation signal on the away-side that is in good agreement with the PHENIX data if we employ the dynamical evolution framework developed in \cite{Mach}.
The most significant contribution to the correlation signal at small angles away from the away-side jet's axis stems from a bow-shock contribution emerging from away-side jets centered around midrapity ($|y|<0.5$) whereas the large angle correlation is induced mainly by away-side jets from a wider range of 
rapidities ($|y|<2$). 
We also discussed how the rapidity structure would appear if the medium would not lead to an elongation of the Mach cone along the rapidity 
axes due to the longitudinal boost in momentum space. The correlation signal would in this case be considerably widened and the maximum of
the correlation signal would be shifted to significantly smaller angles. This is what would appear to have to be realized in scenarios in which no hydrodynamical mode is excited (e.g. gluon radiation). Therefore in general scenarios in which the signal does not couple strongly to the longitudinal flow face new challenges.
  
Our findings hence indicate that medium effects on the signal structure along the beam axis are essential in order to explain the observed correlation. 

  \section*{Acknowledgments}

We'd like to thank K.~Eskola,  U.~Heinz, B.~M\"uller,  V.~Ruuskanen, and  I.~Vitev for useful exchanges. This work was financially supported by the Academy of Finland, Project 206024 and by the Department of Energy, DOE grant DE-FG02-96ER40945. J. R. acknowledges support as a Feodor Lynen fellow of the Alexander v. Humboldt foundation and by the Natural Sciences and Engineering Research Council of 
Canada.

% \appendix

% \section{}
% \label{}

\end{document}